\documentstyle[12pt,epsfig]{article}

\begin{document}
\title{Resonance $f_0(1500)$: Is it a scalar glueball~?}
\author{V.V.Anisovich\\
Petersburg Nuclear Physics Institute\\
Gatchina, St.Petersburg 188350, Russia\\
e-mail:anisovic@lnpi.spb.su}
\date{}
\maketitle
\newcommand{\beq}{\begin{equation}}
\newcommand{\eeq}{\end{equation}}

\begin{abstract}
The ratios of partial widths for the decay of a glueball into two
pseudoscalar mesons are calculated under the assumption that the production
of  light quark pairs ($u\bar u,d\bar d,s\bar s$) in soft gluon-\-induced
reactions  goes on within universal symmetry breaking.
Parameter of the violation of
flavour symmetry is fixed by the central hadron
production data in high energy hadron collisions and/or by the ratios of
radiative decay amplitudes $J/\Psi \to \gamma\eta/\gamma\eta'$
and $J/\Psi \to \gamma \phi \phi/\gamma \omega \omega$.
The ratios of coupling constants $glueball\to\pi\pi,K\bar K,
\eta\eta,\eta\eta'$  which are calculated with this parameter
coincide reasonably with those of $f_0(1500)$, supporting an idea
about glueball nature of $f_0(1500)$.
\end{abstract}

Resonance $f_0(1500)$, which has been discovered by Crystal Barrel
Collaboration
in the simultaneous analysis of the reactions $p\bar p$ (at
rest)$\to\pi^0\pi^0\pi^0$ and $\eta\eta\pi^0$, was
immediately considered as a candidate for the lowest scalar
glueball \cite{1}:
\begin{description}
\item[(i)] Its mass is near the value predicted by the lattice-\-QCD
calculations \cite{2};
\item[(ii)] From the point of view of  $q\bar q$-systematics, this
resonance looks as a superfluous one (the detailed discussion can be found,
for example, in refs. \cite{3,4});
\item[(iii)] In the gluon-rich reaction, $J/\Psi\to \gamma +4\pi$, a sharp
peak is seen in the four-pion spectrum,  which is
  compatible in mass and width
 with $f_0(1500)$; later  the
re-\-analysis of MARK-III data revealed its quantum numbers
to be $0^{++}$ \cite{5}.
\end{description}
But additional arguments in favour of the
glueball nature of
$f_0(1500)$ are needed.
Certain scepticism can
be also extended on the arguments (i) and (ii). The matter is that the
present lattice calculations of the glueball masses do not take properly into
account  quark degrees of freedom.
Still, the  mixture of
 glueball and neighbouring $q\bar q$ states can be significant: this
mixture is not  suppressed  in accordance with the rules of
$1/N$-\-expansion \cite{6,7} ($N=N_c=N_f$ is the number of
colours and light flavours). An admixture of the $q\bar
q$-\-components can shift the mass of the real glueball as compared to
the pure gluodynamic calculations.   Moreover, the
latest
lattice-\-QCD result based on the nowadays larger statistics gives
a substantially higher value for the mass of the pure
glueball, $m(0^{++})=1740\pm50$~MeV \cite{8}.

Systematics of the $q\bar q$-mesons requires  four states
with the quantum numbers of the scalar glueball in the mass
region 1000--1600~MeV  \cite{9}:
 two ground states of the $^3P_0q\bar q$-nonet together with
their radial excitations.          There are candidates  for these
states, see refs. \cite{9}--\cite{11}. However, scalar resonances in the
mass region 1100--1300~MeV have rather large widths: a distant location of
the corresponding poles of the amplitude  from the physical region
allows one to be sceptical about their existence.

In ref. \cite{12}  the possibility to test glueball
nature of $f_0(1500)$ by studying its partial widths
has been emphasized. In this paper the
ratios of couplings of $f_0(1500)$ to $\pi\pi,K\bar
K$,$\eta\eta'$ and $\eta\eta'$ have been calculated. The consideration
presented in this paper is
modified, as compared to that of ref. \cite{12}, in two points:
\begin{description}
\item[(i)] The  diagrams for a glueball decay into two pseudoscalar
$q\bar q$-\-mesons are selected according to $1/N$-\-expansion rules;
the leading diagrams are taken into account only. This reduces the
number of unknown parameters.
\item[(ii)] The flavour-symmetry violation parameter, $\lambda$,
is fixed by  the ratio of strange and non-\-strange hadrons
produced in the central
region of high energy hadron collisions and/or by the branching ratios of
the radiative decays
$J/\Psi\to\gamma\eta/\gamma\eta'$
and $J/\Psi \to \gamma \phi \phi/\gamma \omega \omega$.
\end{description}
This gives explicit values for the ratios of coupling constants
$glueball\to\pi\pi,K\bar K,\eta\eta,\eta\eta'$ which coincide rather
reasonably  with those of $f_0(1500)$.

Two types of processes determine the glueball decay into two pseudoscalar
mesons \cite{12,13}. They are shown in Figs. 1a and 1b or more
schematically  in Figs. 1c and 1d. According to
$1/N$-\-expansion rules, the main contribution is given by the diagram of
Fig.1c, while  the contribution of the process of Fig. 1d  is
suppressed by the
factor $1/N$. Indeed,  the coupling constants
which determine the  process of
Fig.1c have the following order of magnitudes:
$g_{GL}$($glueball\to q\bar q)\sim1/N$ and $g_m$($meson\to q\bar
q)\sim1/\sqrt{N}$. So, the vertex of Fig.1c is of the order of
$g_{GL}\,g^2_m\,N_c\sim 1/N$. The vertex of the diagram of the Fig.1d is
determined by additional couplings $\widetilde g_{GL}$($glueball\to
gg)\sim 1/N$ and $g(g\to q\bar q)\sim1/\sqrt{N}$, so it is of the order
of $\widetilde g_{GL}^4\,g^2_m\,N^2_c\sim1/N^2$. Let us note that  quark
content of the produced mesons in the diagrams  of Figs.1c and 1d is
fixed; thus,  quark loops in these diagrams do not generate the
enlarging factor $N_f$ \cite{7}. Below the contribution  of the
process of Fig.1c is taken into account only; the next-\-to-\-leading
contribution (Fig.1d) is neglected.

In the process of Fig.1c (or Fig.1a), two
$q\bar q$-pairs are produced by soft gluons, and this production
 goes with flavour symmetry violation.  Clear indication
on this violation is seen in other gluon-\-rich
processes. Central production of hadrons in the hadron collisions at high
energies is determined by the following process: the gluons of the pomeron
create $q\bar q$-pairs  which are then transformed into hadrons.
Fig. 2a provides an example of a diagram
 which describes the $q\bar q$ production cross section.
Experimental data at high and superhigh energies confirm the suppresion
of the $s\bar s$-production:
\beq
u\bar u:d\bar d:s\bar s=1:1:\lambda.
\eeq

The production probability factor, $\lambda$, is related to the
directly produced
hadrons (i.e. to the hadrons which are not resonance decay product); it is
equal to $\lambda=0.4-0.5$ (see \cite{14} and references therein).

Branching ratios for the radiative decays $J/\Psi\to\gamma\eta$
and $\gamma\eta'$ give us another evidence for the flavour symmetry
breaking with the same parameter $\lambda$ as in the central hadron
production. Hadron production in the radiative $J/\Psi$-decay is
determined by the $c\bar c$-annihilation into two gluons \cite{15} with the
subsequent production of the light-\-quark $q\bar q$-pair (see Fig. 2b).
The process of Fig. 2b leads to the following ratio for the decay couplings:
\beq
\frac{g(J/\Psi\to\gamma\eta')}{g(J/\Psi\to\gamma\eta)}=\frac{\sqrt{2}
\sin\Theta+\sqrt{\lambda}\cos\Theta}{\sqrt{2}\cos\Theta
-\sqrt{\lambda}\sin\Theta}.
\eeq
Here $\Theta$ is the mixing angle for $n\bar n=(u\bar u+d\bar d)/\sqrt{2}$
and $s\bar s$ components in $\eta$ and $\eta'$, namely, $\eta=\cos\Theta
n\bar n-\sin\Theta s\bar s$ and $\eta'=\sin \Theta n\bar n+\cos\Theta s\bar
s$. The branching ratios $\gamma\eta$ and $\gamma\eta'$ are equal to
$B(\gamma\eta)=(8.6\pm0.8)\cdot10^{-4}$ and $B(\gamma\eta')
=(4.3\pm0.3)\cdot10^{-3}$ \cite{9}, so the ratio of the decay
amplitudes, Eq.(2), is equal to $2.02\pm0.23$. This gives
$\lambda=0.48\pm0.09$.

The processes
 $J/\Psi \to \gamma \phi \phi$ and $J/\Psi \to \gamma \omega \omega$
                 present  non-resonance production of two
$q\bar q$-mesons. They are shown in Fig. 2c. Supposing "ideal"
$\phi$ -- $\omega$ mixing , $\phi=s\bar s$ and $\omega=n\bar n$, one has
the following ratio for the decay couplings:
\beq
\frac{g(J/\Psi\to\gamma\phi \phi)}{g(J/\Psi\to\gamma\omega \omega)}=
\lambda.
\eeq
Corresponding branching ratios, $B(\gamma \phi \phi)=(4.0 \pm1.2)
10^{-4}$ and  $B(\gamma \omega \omega)=(15.9 \pm3.3)10^{-4}$ [9], give
$\lambda\simeq 0.5$ again. Thus, in the radiative $J/\Psi$-decays the
strange-quark production probability factor, $\lambda$, is the same
as in the high-energy central production,
supporting an idea of universality of the parameter
$\lambda$ in soft gluonic processes.

Now let us return to the discussion of the glueball decay into two
pseudoscalar mesons. The decay coupling constants, being determined
by the process of Fig.1c, differ only in factors which depend on
the quark content of  pseudoscalar mesons and the suppression
parameter $\lambda$. These factors squared are presented in Table
1. As a result, the partial width ratios, corrected in phase spaces
$\gamma^2=\Gamma/\phi$, are equal to
$$ \gamma^2(\pi\pi):\gamma^2(K\bar
K):\gamma^2(\eta\eta):\gamma^2(\eta\eta'):\gamma^2(\eta'\eta')=$$
\beq
1:\frac43\,\lambda
:\frac13(\cos^2\Theta+\lambda\sin^2\Theta)^2:\frac23(1-\lambda)^2\sin^2
\Theta\cos^2\Theta:\frac13(\sin^2\Theta+\lambda\cos^2\Theta)^2.
\eeq
With $\Theta=37^\circ$ \cite{9} and $\lambda=0.4$, we have for the
righthand side of Eq. (3):
\beq
1:0.53:0.20:0.06:0.13.
\eeq
These ratios are in a reasonable agreement with nowadays partial
width data for $f_0(1500)$ obtained in the $p\bar p$-annihilation at
rest: $\gamma^2(K\bar K)/\gamma^2(\pi\pi)=0.49\pm0.16$ \cite{16};
$\gamma^2(\eta\eta)/\gamma^2(\pi\pi)=0.27\pm0.11$ \cite{12},
$0.23\pm0.09$ \cite{17}; $\gamma^2(\eta\eta')/\gamma^2(\pi\pi)=
0.19\pm0.08$ \cite{12}, $0.33\pm0.16$ \cite{17}. This allows us to
consider $f_0(1500)$ as a good candidate for glueball. However,  more
precise extraction of branching ratios is definitely required.
The main problem in finding partial widths is an interference
of the $f_0(1500)$ signal with the background, which causes additional
systematic errors. For example,  another
parametrization of the background \cite{18} leads to the ratio
$\gamma^2(\eta\eta)/\gamma^2(\pi\pi)$ twice as small as that of
ref. \cite{12}. This emphasizes a necessity of further study of the
$f_0(1500)$ branching ratios, especially with other possible
backgrounds, namely, in different reactions or at different $p\bar
p$-energies.

In ref. [19] the glueball origin of $f_J(1710)$ has been supposed.
However, the measured ratio of the decay probabilities $f_J(1710)\to \pi\pi/
K\bar K$ is of the order of 0.3 [9,20] which disagrees strongly with the idea
of suppression of the strange quark production in  gluonic processes
(see Table 1 for $\pi \pi/ K\bar K$ ratio).

The factors given in Table 1, which describe the transition $gluons \to two\;
q\bar q\; mesons$, can be applied not only to the glueball decay.
As was mentioned above, the same
factors give relative probabilities for the non-resonance production
of the two $q\bar q$-mesons in the radiative $J/\Psi$-decay (see Fig. 2c).
Therefore, these  factors can  be used as a test of the gluonic nature of
resonances produced in the radiative $J/\Psi$-decay, because
the glueball decay
products as well as the background events obey the relations given in
Eq. (4), while the production via $q\bar q$-resonance should disturb these
relations.

In conclusion, the high energy central production data give a clear
evidence that light quarks in soft gluonic processes are produced
with the  flavour symmetry violation: the strange quark production
is suppressed. The radiative decay ratios
$J/\Psi\to\gamma\eta/\gamma\eta'$ and $\gamma \phi \phi/\gamma
\omega \omega$ reveal the same suppression factor
in the transition $gluons\to q\bar q$ as the high energy production data.
Coupling constants for the glueball decay into two $q\bar q$-mesons
are calculated using the idea of universality of the flavour symmetry
violation in the production of soft $q\bar q$-pairs by gluons. These coupling
constants are in reasonable agreement with the Crystal Barrel
 data for the decay ratios of $f_0(1500)$ into two
pseudoscalar mesons, thus  arguing in favour of glueball nature
of $f_0(1500)$. The obtained ratios can be also
used in the radiative $J/\Psi$-decays to test the glueball origin
of the produced resonances.

I am grateful to D.V.Bugg, F.E.Close, L.G.Dakhno, S.S.Gerstein, L.Montanet,\\
A.V.Sarantsev, and B.S.Zou for useful discussions and
helpful remarks. I acknowledge
financial support from the Royal Society during my visit to QMWC,
London University, where this work was partly done. I thank
J.-P.Stroot for kind hospitality at "Gluonium~95", where a preliminary
version of this paper has been presented.

\newpage

{\bf Table 1.} Coupling constants squared for the  glueball decay\\

\begin{center}
\begin{tabular}{||c|c|c||} \hline\hline
Decay & Coupling & Identity \\ && particle \\
channel & squared & factor\\ \hline \\ && \\
$\pi^0\pi^0$ & 1 & 1/2 \\ && \\
$\pi^+\pi^-$ & 1& 1\\ &&\\
$K^+K^-$ & $\lambda$ & 1 \\ && \\
$K^0\bar K^0$ & $\lambda$ & 1 \\ && \\
$\eta\eta$ & $(\cos^2\Theta+\lambda\sin^2\Theta)^2$ & 1/2 \\ &&\\
$\eta\eta'$ & $(1-\lambda)^2\sin^2\Theta\cos^2\Theta$ & 1\\ &&\\
$\eta'\eta'$ & $(\sin^2\Theta+\lambda\cos^2\Theta)^2$ & 1/2\\
\hline\hline
\end{tabular} \end{center}

\newpage

\begin{figure}
\begin{center}
\mbox{\epsfig{file=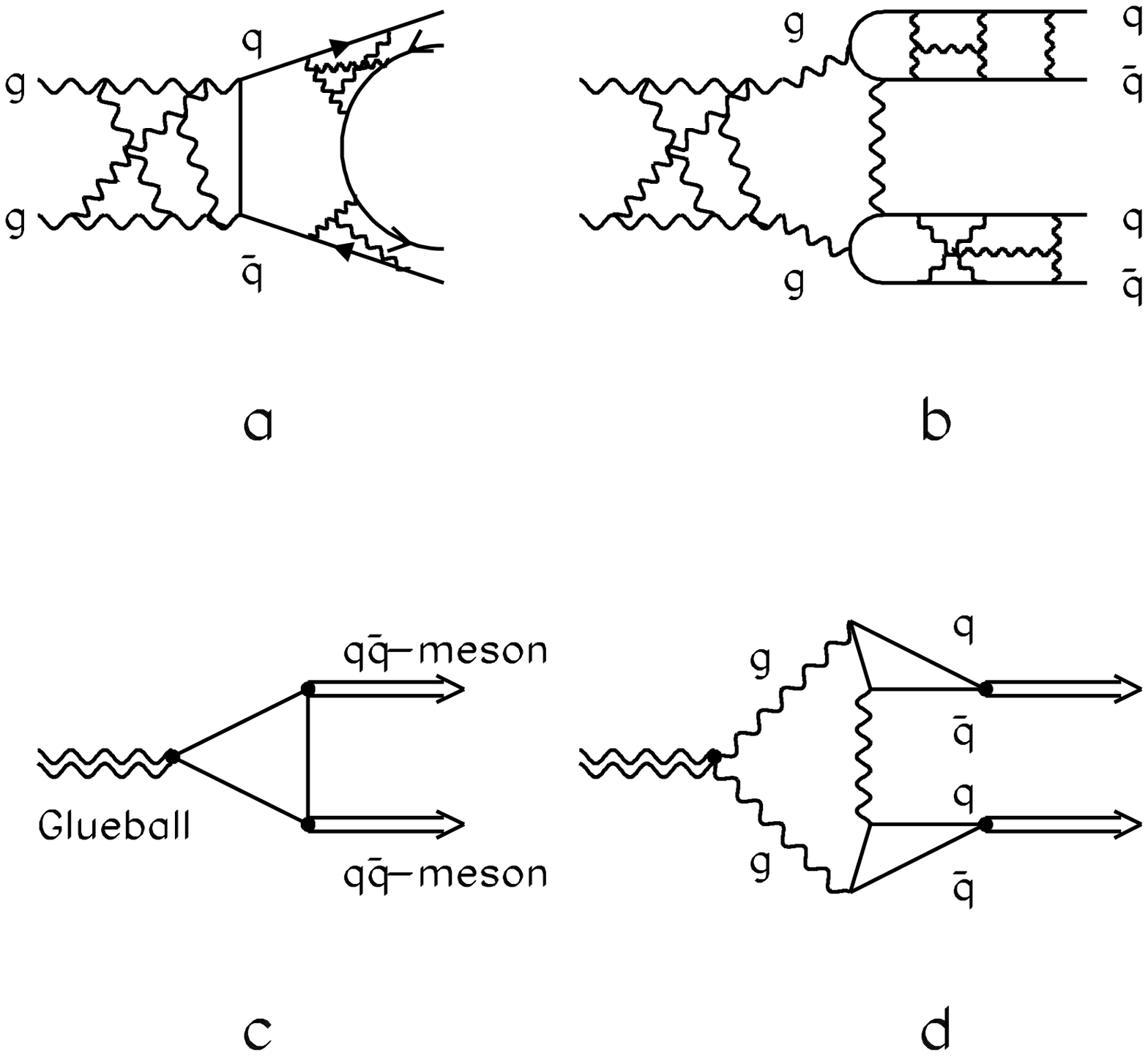,width=12cm}}
\end{center}
Fig. 1.  Production of $q\bar q$-pair in the decay
$glueball \to two\;q\bar q \;mesons$.
\end{figure}

\newpage

\begin{figure}
\begin{center}
\mbox{\epsfig{file=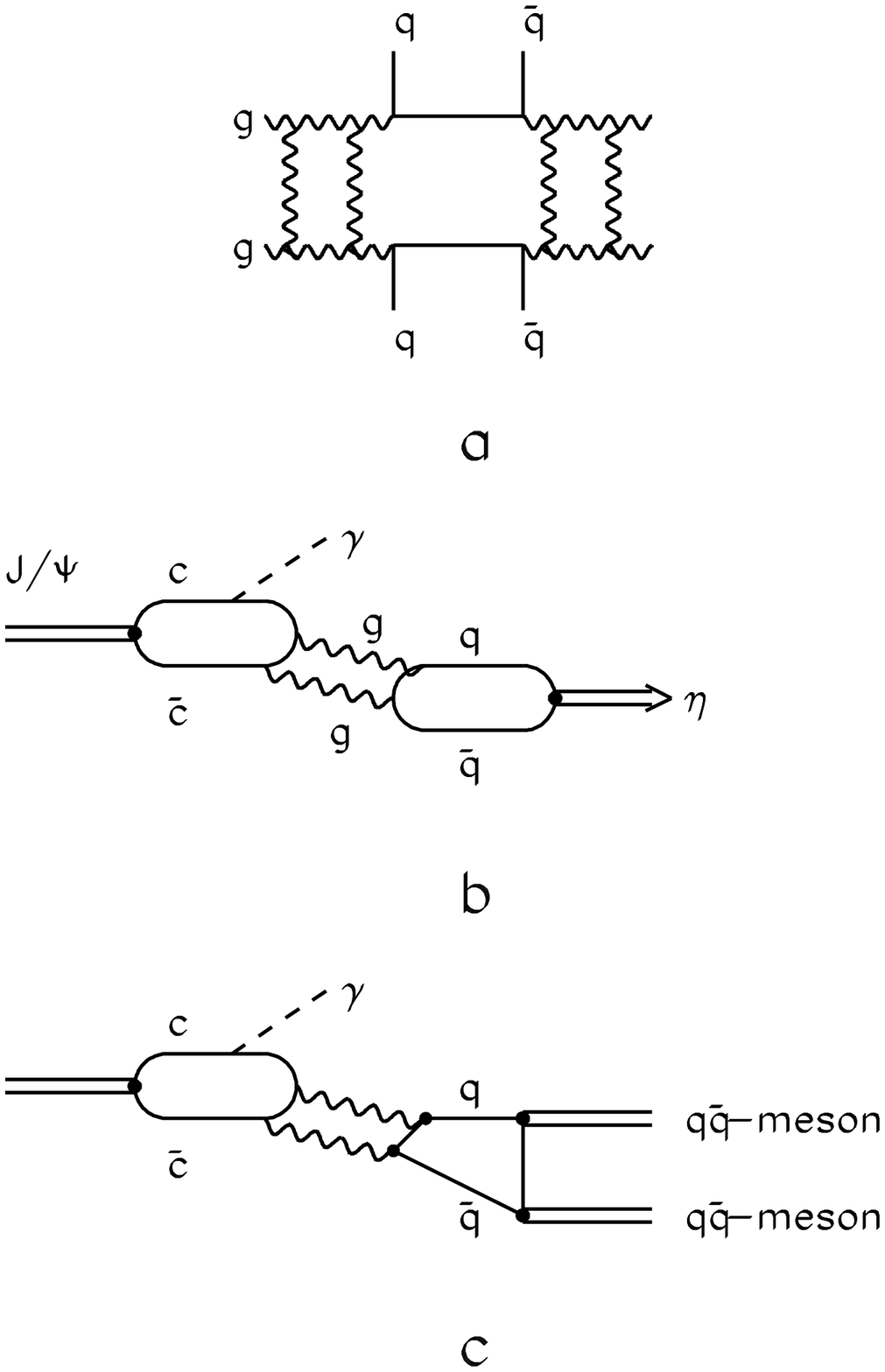,width=15cm}}
\end{center}
Fig. 2.  Production of $q\bar q$-pairs in the gluon ladder of
the pomeron (a), and in the radiative $J/\Psi$-decay (b,c).
\end{figure}

\end{document}